\documentclass{iopart}
\usepackage{amssymb}
\usepackage{graphicx}
\usepackage{comment}

\begin{document}

\title{Fundamental parameter-free solutions in Modified Gravity}

\author{J. W. Moffat$^{1,2}$ and V. T. Toth$^1$\\
{\rm
\footnotesize
$^1$Perimeter Institute for Theoretical Physics, Waterloo, Ontario N2L 2Y5, Canada\\
$^2$Department of Physics, University of Waterloo, Waterloo, Ontario N2L 3G1, Canada}}

\begin{abstract}
Modified Gravity (MOG) has been used successfully to explain the rotation curves of galaxies, the motion of galaxy clusters, the Bullet Cluster, and cosmological observations without the use of dark matter or Einstein's cosmological constant. We now have the ability to demonstrate how these solutions can be obtained directly from the action principle, without resorting to the use of fitted parameters or empirical formulae. We obtain numerical solutions to the theory's field equations that are exact in the sense that no terms are omitted, in two important cases: the spherically symmetric, static vacuum solution and the cosmological case of an homogeneous, isotropic universe. We compare these results to selected astrophysical and cosmological observations.
\end{abstract}


\maketitle

\section{Introduction}

Our modified gravity (MOG) theory, also known in previous work \cite{Moffat2006a} as Scalar-Tensor-Vector Gravity or STVG, is based on an action that incorporates, in addition to the Einstein-Hilbert term and the matter action, a massive vector field, and three scalar fields corresponding to running values of the gravitational constant, the vector field coupling constant, and the vector field mass.

A Dirac-Hamiltonian analysis \cite{IN1977} can be used to show that the theory is free of constraints and it is free of ghosts and instabilities. The theory has been used successfully to explain cosmological observations \cite{Moffat2007c}, the motion of galaxy clusters \cite{Brownstein2006b}, the Bullet Cluster \cite{Brownstein2007}, the rotation curves of galaxies and dwarf galaxies \cite{Brownstein2006a}, and the velocity dispersion profiles of satellite galaxies \cite{Moffat2007,Moffat2009a} and globular clusters \cite{Moffat2007a} without exotic dark matter. However, until now these explanations required case-by-case fitting of two parameters, or alternatively, {\em ad-hoc} formulae, not derivable from the action principle, to predict the values of these parameters.

In this paper, we demonstrate that it is in fact possible to eliminate fitted parameters altogether in favor of universal constants. After fixing initial conditions using observational data, the field equations can be integrated numerically, and the theory can be used to make predictions from the scale of the solar system to cosmological scales, i.e., across at least 14 orders of magnitude in length, or more than 20 orders of magnitude in mass-energy.

We begin in Section~\ref{sec:theory} by introducing the theory through the action principle, and establish key assumptions that allow us to analyze physically relevant scenarios. In Section~\ref{sec:fldeqs}, we derive the field equations using the variational principle. In Section~\ref{sec:static} we solve the field equations in the static, spherically symmetric case. In Section~\ref{sec:testpart}, we postulate the action for a test particle, and obtain approximate solutions to the field equations for a spherically symmetric gravitational field. In Section~\ref{sec:cosmo} we demonstrate how the Friedmann equations of cosmology can be obtained from the theory. In Section~\ref{sec:obs}, we utilize the theory to obtain new estimates for galaxy rotation curves, derive the Tully-Fisher law, and show how the solutions we obtained for the field equations remain valid from cosmological to solar system scales. Lastly, we end in Section~\ref{sec:concl} with conclusions.

\section{Modified Gravity Theory}
\label{sec:theory}

The action of our theory is constructed as follows \cite{Moffat2006a}. We start with the Einstein-Hilbert Lagrangian density that describes the geometry of spacetime:
\begin{equation}
{\cal L}_G=-\frac{1}{16\pi G}\left(R+2\Lambda\right)\sqrt{-g},
\end{equation}
where $G$ is the gravitational constant, $g$ is the determinant of the metric tensor $g_{\mu\nu}$ (we are using the metric signature $(+,-,-,-)$), and $\Lambda$ is the cosmological constant. We set the speed of light, $c=1$. The Ricci-tensor is defined as
\begin{equation}
R_{\mu\nu}=\partial_\alpha\Gamma^\alpha_{\mu\nu}-\partial_\nu\Gamma^\alpha_{\mu\alpha}+\Gamma^\alpha_{\mu\nu}\Gamma^\beta_{\alpha\beta}-\Gamma^\alpha_{\mu\beta}\Gamma^\beta_{\alpha\nu},
\end{equation}
where $\Gamma^\alpha_{\mu\nu}$ is the Christoffel-symbol, while $R=g^{\mu\nu}R_{\mu\nu}$.

We introduce a ``fifth force'' vector field $\phi_\mu$ via the Maxwell-Proca Lagrangian density:
\begin{equation}
{\cal L}_\phi=-\frac{1}{4\pi}\omega\left[\frac{1}{4}B^{\mu\nu}B_{\mu\nu}-\frac{1}{2}\mu^2\phi_\mu\phi^\mu+V_\phi(\phi)\right]\sqrt{-g},
\end{equation}
where $B_{\mu\nu}=\partial_\mu\phi_\nu-\partial_\nu\phi_\mu$, $\mu$ is the mass of the vector field, $\omega$ characterizes the strength of the coupling between the ``fifth force'' and matter, and $V_\phi$ is a self-interaction potential.

Next, we promote the three constants of the theory, $G$, $\mu$ and $\omega$, to scalar fields by introducing associated kinetic and potential terms in the Lagrangian density:
\begin{eqnarray}
{\cal L}_S&=&-\frac{1}{G}\left[\frac{1}{2}g^{\mu\nu}\left(\frac{\nabla_\mu G\nabla_\nu G}{G^2}+\frac{\nabla_\mu\mu\nabla_\nu\mu}{\mu^2}-\nabla_\mu\omega\nabla_\nu\omega\right)\right.\nonumber\\
&&\left.+\frac{V_G(G)}{G^2}+\frac{V_\mu(\mu)}{\mu^2}+V_\omega(\omega)\right]\sqrt{-g},
\end{eqnarray}
where $\nabla_\mu$ denotes covariant differentiation with respect to the metric $g_{\mu\nu}$, while $V_G$, $V_\mu$, and $V_\omega$ are the self-interaction potentials associated with the scalar fields.

Our action integral takes the form
\begin{equation}
S=\int{({\cal L}_G+{\cal L}_\phi+{\cal L}_S+{\cal L}_M)}~d^4x,
\label{eq:FldL}
\end{equation}
where ${\cal L}_M$ is the ordinary matter Lagrangian density, such that the energy-momentum tensor of matter takes the form:
\begin{equation}
T_{\mu\nu}=-\frac{2}{\sqrt{-g}}\frac{\delta S_M}{\delta g^{\mu\nu}},
\end{equation}
where $S_M=\int{\cal L}_M~d^4x$. A ``fifth force'' matter current can be defined as:
\begin{equation}
J^\nu=-\frac{1}{\sqrt{-g}}\frac{\delta S_M}{\delta\phi_\nu}.
\end{equation}

We assume that the variation of the matter action with respect to the scalar fields vanishes:
\begin{equation}
\frac{\delta S_M}{\delta X}=0,
\end{equation}
where $X=G,\mu,\omega$.

\section{Field equations}
\label{sec:fldeqs}

The field equations of the theory can be obtained in the form of the first and second-order Euler-Lagrange equations:
\begin{eqnarray}
\frac{\partial{\cal L}}{\partial X}-\nabla_\nu\frac{\partial{\cal L}}{\partial(\partial_\nu X)}&=&0,\\
\frac{\partial{\cal L}}{\partial \phi_\mu}-\nabla_\nu\frac{\partial{\cal L}}{\partial(\nabla_\nu\phi_\mu)}&=&J^\mu,\\
\frac{\partial{\cal L}}{\partial g^{\mu\nu}}-\partial_\kappa\frac{\partial{\cal L}}{\partial g^{\mu\nu}{}_{,\kappa}}+\partial_\kappa\partial_\lambda\frac{\partial{\cal L}}{\partial g^{\mu\nu}{}_{,\kappa\lambda}}&=&T_{\mu\nu},
\end{eqnarray}
where ${\cal L}={\cal L}_G+{\cal L}_\phi+{\cal L}_S+{\cal L}_M$ is the total Lagrangian density and a comma in the covariant index is used to indicate a coordinate derivative.

The full set of the theory's field equations reads:
\begin{equation}
\frac{1}{4\pi}\left[\omega\nabla_\mu B^{\mu\nu}+\nabla_\mu\omega B^{\mu\nu}+\omega\mu^2\phi^\nu-\omega\frac{\partial V_\phi(\phi)}{\partial\phi_\nu}\right]=J^\nu,
\end{equation}
\begin{equation}
\nabla^\nu\nabla_\nu\mu-\frac{\nabla^\nu\mu\nabla_\nu\mu}{\mu}-\frac{\nabla^\nu G\nabla_\nu\mu}{G}+\frac{1}{4\pi}G\omega\mu^3\phi_\mu\phi^\mu+\frac{2}{\mu}V_\mu(\mu)-V'_\mu(\mu)=0,
\end{equation}
\begin{eqnarray}
\nabla^\nu\nabla_\nu\omega-\frac{\nabla^\nu G\nabla_\nu\omega}{G}-\frac{1}{8\pi}G\mu^2\phi_\mu\phi^\mu+\frac{G}{16\pi}B^{\mu\nu}B_{\mu\nu}+\frac{1}{4\pi}GV_\phi(\phi)
\nonumber\\
+V'_\omega(\omega)=0,
\end{eqnarray}
\begin{eqnarray}
\nabla^\nu\nabla_\nu G-\frac{3}{2}\frac{\nabla^\nu G\nabla_\nu G}{G}+\frac{G}{2}\left(\frac{\nabla^\nu\mu\nabla_\nu\mu}{\mu^2}-\nabla^\nu\omega\nabla_\nu\omega\right)
+\frac{3}{G}V_G(G)\nonumber\\
-V'_G(G)+G\left[\frac{V_\mu(\mu)}{\mu^2}+V_\omega(\omega)\right]+\frac{G}{16\pi}(R+2\Lambda)=0,
\end{eqnarray}
\begin{eqnarray}
\left(\frac{2\nabla_\alpha G\nabla_\beta G}{G^2}-\frac{\nabla_\alpha\nabla_\beta G}{G}\right)(g^{\alpha\beta}g_{\mu\nu}-\delta^\alpha_\mu\delta^\beta_\nu)
\nonumber\\
-8\pi\left[\left(\frac{1}{4\pi}G\omega\mu^2\phi_\alpha\phi_\beta-\frac{\partial_\alpha G\partial_\beta G}{G^2}-\frac{\partial_\alpha\mu\partial_\beta\mu}{\mu^2}+\partial_\alpha\omega\partial_\beta\omega\right)\right.
\nonumber\\
\times\left(\delta^\alpha_\mu\delta^\beta_\nu-\frac{1}{2}g^{\alpha\beta}g_{\mu\nu}\right)
\nonumber\\
\left.+\frac{1}{4\pi}G\omega\left(B^\alpha{}_\mu B_{\nu\alpha}+\frac{1}{4}g_{\mu\nu}B^{\alpha\beta}B_{\alpha\beta}\right)\right.
\nonumber\\
\left.+g_{\mu\nu}\left(\frac{1}{4\pi}GV_\phi(\phi)+\frac{V_G(G)}{G^2}+\frac{V_\mu(\mu)}{\mu^2}+V_\omega(\omega)\right)\right]
\nonumber\\
+R_{\mu\nu}-\frac{1}{2}g_{\mu\nu}R+g_{\mu\nu}\Lambda
=-8\pi GT_{\mu\nu}.
\end{eqnarray}

\section{Static, Spherically Symmetric Vacuum Solution}
\label{sec:static}

In the static, spherically symmetric case with line element
\begin{equation}
ds^2=Bdt^2-Adr^2-r^2d\Omega^2,
\end{equation}
with $d\Omega^2=d\theta^2+\sin^2{\theta}d\phi^2$, the field equations are written as
\begin{equation}
\frac{1}{A}\mu^2\phi_r+\frac{\partial V_\phi}{\partial\phi_r}=\frac{4\pi}{A\omega}J_r,
\end{equation}
\begin{eqnarray}
\phi_t''+\frac{2}{r}\phi_t'+\frac{\omega'}{\omega}\phi_t'+\frac{1}{2}\left(3\frac{A'}{A}-\frac{B'}{B}\right)\phi_t'
-A\mu^2\phi_t+AB\frac{\partial V_\phi}{\partial\phi_t}
\nonumber\\
=-\frac{4\pi A}{\omega}J_t,
\label{eq:phi}
\end{eqnarray}
\begin{eqnarray}
G''+\frac{2}{r}G'-\frac{3}{2}\frac{G'^2}{G}+\frac{1}{2}\left(\frac{\mu'^2}{\mu^2}-\omega'^2\right)G+\frac{1}{2}\left(\frac{B'}{B}-\frac{A'}{A}\right)G'&&\nonumber\\
+AV'_G(G)-3A\frac{V_G(G)}{G}-AG\left[\frac{V_\mu(\mu)}{\mu^2}+V_\omega(\omega)\right]-\frac{AG(R+2\Lambda)}{16\pi}
\nonumber\\
=0,
\end{eqnarray}
\begin{eqnarray}
\mu''+\frac{2}{r}\mu'-\frac{\mu'^2}{\mu}-\frac{G'}{G}\mu'+\frac{1}{4\pi}G\omega\left(\phi_r^2-\frac{A}{B}\phi_t^2\right)\mu^3
\nonumber\\
+\frac{1}{2}\left(\frac{B'}{B}-\frac{A'}{A}\right)\mu'-2A\frac{V_\mu(\mu)}{\mu}+AV'_\mu(\mu)=0,
\end{eqnarray}
\begin{eqnarray}
\omega''+\frac{2}{r}\omega'-\frac{G'}{G}\omega'+\frac{1}{8\pi}G\mu^2\left(\frac{A}{B}\phi_t^2-\phi_r^2\right)+\frac{1}{2}\left(\frac{B'}{B}-\frac{A'}{A}\right)\omega'
\nonumber\\
+\frac{1}{8\pi B}G\phi_t'^2-\frac{1}{4\pi}AGV_\phi(\phi)-AV'_\omega(\omega)=0,
\end{eqnarray}
\begin{eqnarray}
8\pi GT^t_t=-\Lambda-V-\frac{1}{A}N+\frac{A'}{A^2r}-\frac{1}{Ar^2}+\frac{1}{r^2}+\frac{G''}{AG}+\frac{2}{r}\frac{G'}{AG}
\nonumber\\
-2\frac{G'^2}{AG^2}-\frac{1}{2}\frac{A'G'}{A^2G}-\omega G\left(\frac{\phi_t'^2}{AB}+\frac{\mu^2\phi_t^2}{B}+\frac{\mu^2\phi_r^2}{A}\right),
\end{eqnarray}
\begin{eqnarray}
8\pi GT^r_r=-\Lambda-V+\frac{1}{A}N-\frac{B'}{ABr}-\frac{1}{Ar^2}+\frac{1}{r^2}+\frac{1}{2}\frac{B'G'}{ABG}+\frac{2}{r}\frac{G'}{AG}
\nonumber\\
-\omega G\left(\frac{\phi_t'^2}{AB}-\frac{\mu^2\phi_t^2}{B}-\frac{\mu^2\phi_r^2}{A}\right),
\end{eqnarray}
\begin{equation}
8\pi GT^t_r=-2\frac{G\omega\mu^2\phi_t\phi_r}{B},~~~~~~~~~8\pi GT^r_t=2\frac{G\omega\mu^2\phi_t\phi_r}{A},
\end{equation}
\begin{eqnarray}
8\pi GT^\theta_\theta=8\pi GT^\phi_\phi=-\Lambda-V-\frac{1}{A}N
+\frac{1}{2}\frac{A'}{A^2r}+\frac{1}{4}\frac{A'B'}{A^2B}
\nonumber\\
-\frac{1}{2}\frac{B'}{ABr}+\frac{1}{4}\frac{B'^2}{AB^2}
-\frac{1}{2}\frac{B''}{AB}+\frac{G''}{AG}+\frac{1}{r}\frac{G'}{AG}-2\frac{G'^2}{AG^2}-\frac{1}{2}\frac{A'G'}{A^2G}
\nonumber\\
+\frac{1}{2}\frac{B'G'}{ABG}
-\omega G\left(-\frac{\phi_t'^2}{AB}-\frac{\mu^2\phi_t^2}{B}+\frac{\mu^2\phi_r^2}{A}\right),
\end{eqnarray}
where
\begin{equation}
R=\frac{B''}{AB}-\frac{B'^2}{2AB^2}-\frac{A'B'}{2A^2B}+\frac{2B'}{ABr}-\frac{2A'}{A^2r}+\frac{2}{Ar^2}-\frac{2}{r^2},
\end{equation}
\begin{equation}
N=-4\pi\left(\frac{\mu'^2}{\mu^2}+\frac{G'^2}{G^2}-\omega'^2\right),
\end{equation}
\begin{equation}
V=2\omega GV_\phi(\phi)+8\pi\left[\frac{V_G(G)}{G^2}+\frac{V_\mu(\mu)}{\mu^2}+V_\omega(\omega)\right].
\end{equation}
The prime denotes differentiation with respect to $r$, i.e., $y'=dy/dr.$

These equations can be substantially simplified in the matter vacuum case ($T^\mu_\nu=0$), with no cosmological constant ($\Lambda=0$), setting the potentials to zero ($V_\phi=V_G=V_\mu=V_\omega=0$) and also setting $\phi_r=0$. These choices leave us with six equations in the six unknown functions $A$, $B$, $\phi_t$, $G$, $\mu$, and $\omega$, which read after some trivial rearranging:
\begin{eqnarray}
\frac{B'G'}{2ABG}-\frac{G'}{AGr}+2\omega G\left(\frac{\phi_t'^2}{AB}+\frac{\mu^2\phi_t^2}{B}\right)-\frac{B''}{2AB}+\frac{B'^2}{4AB^2}
\nonumber\\
+\frac{A'B'}{4A^2B}-\frac{B'}{2ABr}-\frac{A'}{2A^2r}+\frac{1}{Ar^2}-\frac{1}{r^2}=0,
\end{eqnarray}
\begin{eqnarray}
\frac{G''}{AG}-\frac{2G'^2}{AG^2}-\frac{B'G'}{2ABG}-\frac{A'G'}{2A^2G}+\frac{B'}{ABr}+\frac{A'}{A^2r}
\nonumber\\
+8\pi\left(\frac{G'^2}{AG^2}+\frac{\mu'^2}{A\mu^2}-\frac{\omega'^2}{A}-\frac{\omega G\mu^2\phi_t^2}{4\pi B}\right)=0,
\end{eqnarray}
\begin{eqnarray}
\omega G\left(\frac{\phi_t'^2}{AB}-\frac{\mu^2\phi_t^2}{B}\right)+4\pi\left(\frac{G'^2}{AG^2}+\frac{\mu'^2}{A\mu^2}-\frac{\omega'^2}{A}\right)
\nonumber\\
+\frac{B'G'}{2ABG}+\frac{2G'}{AGr}-\frac{B'}{ABr}-\frac{1}{Ar^2}+\frac{1}{r^2}=0,
\end{eqnarray}
\begin{equation}
\mu''+\frac{2}{r}\mu'-\frac{\mu'^2}{\mu}-\frac{G'}{G}\mu'+\frac{1}{2}\left(\frac{B'}{B}-\frac{A'}{A}\right)\mu'-\frac{A\omega G\phi_t^2}{4\pi B}\mu^3=0,\label{eq:mu}
\end{equation}
\begin{equation}
\omega''+\frac{2}{r}\omega'-\frac{G'}{G}\omega'+\frac{1}{2}\left(\frac{B'}{B}-\frac{A'}{A}\right)\omega'+\frac{G}{2B}\phi_t'^2+\frac{AG\mu^2\phi_t^2}{8\pi B}=0,\label{eq:omega}
\end{equation}
\begin{eqnarray}
G''+\frac{2}{r}G'-\frac{3}{2}\frac{G'^2}{G}+\frac{1}{2}\left(\frac{B'}{B}-\frac{A'}{A}\right)G'+\frac{1}{2}\left(\frac{\mu'^2}{\mu^2}-\omega'^2\right)G
\nonumber\\
-\frac{1}{16\pi}AGR=0,\label{eq:G}
\end{eqnarray}

A solution to this set of six equations can be obtained using numerical methods. This solution is exact in the sense that it is obtained without dropping any terms, and its accuracy is limited only by the machine precision and the stability of the integration algorithm. This approach requires that we find suitable initial conditions for the six unknown functions $A$, $B$, $G$, $\mu$, $\omega$, and $\phi_t$. This is accomplished in the next section, using the motion of a point test particle as a guide.

\begin{figure}[t]
\hskip 0.2\linewidth\includegraphics[width=0.8\linewidth]{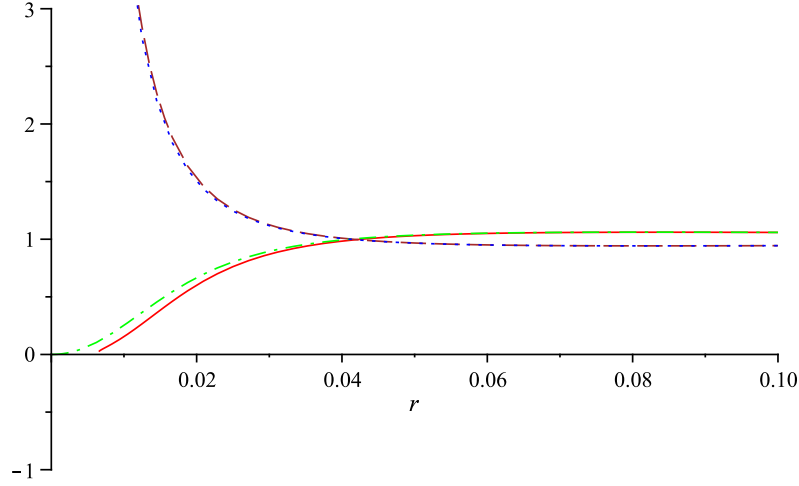}
\caption{Comparing MOG numerical solutions to the Reissner-Nordstr\"om solution, for a $10^{11}~M_\odot$ source mass. The MOG parameters $A$ (solid red line) and $B$ (dashed brown line) are plotted along with the Reissner-Nordstr\"om values of $A$ (dash-dot green line) and $B$ (dotted blue line), calculated using (\ref{eq:BB}). Horizontal axis is in pc. We observe that the $A$ parameter reaches 0 at below the Schwarzschild radius of a $10^{11}~M_\odot$ mass, which is $\sim 0.01$~pc.}
\label{fig:AB}
\end{figure}

Before we proceed to the case of the test particle, however, let us describe in approximate terms the type of solution that we obtain through numerical methods. The following findings apply when the distance $r$ from the source mass is much larger than the Schwarzschild radius $r_S=2GM$ associated with the source mass, $r\gg r_S$:
\begin{itemize}
\item All three scalar fields turn out to be constant functions, not dependent on $r$:
\begin{eqnarray}
G(r)&\simeq&G_0,\\
\mu(r)&\simeq&\mu_0,\\
\omega(r)&\simeq&\omega_0.
\end{eqnarray}
\item The timelike component $\phi_t$ of the vector field has a negative exponential relationship to the radial distance:
\begin{equation}
\phi_t\simeq-Q_5\frac{e^{-\mu r}}{r},
\label{eq:phit}
\end{equation}
where $Q_5$ acts as a fifth-force charge. Note that in the large $r$ limit, this result can also be obtained directly from Eq.~(\ref{eq:phi}), after setting $A'=B'=\omega'=V_\phi(\phi)=J_t=0$ and dropping the nonphysical positive exponential solution.
\item The metric parameters $A$ and $B$ are in very close agreement with a Reissner-Nordstr\"om type solution for the spherically symmetric field of a charged mass:
\begin{eqnarray}
B(r)&\simeq&1-\frac{2G_NM}{r}+\frac{\omega G_0Q_5^2}{r^2},\label{eq:B}\\
A(r)&\simeq&B(r)^{-1},
\end{eqnarray}
where $G_N$ is Newton's gravitational constant, $Q_5$ is the ``fifth force'' charge which is proportional to the source mass $M$. The emergence of the Reissner-Nordstr\"om solution, despite the fact that unlike electromagnetism, our vector field is massive, can be explained by the fact that at small values of $r$, mass contributions in the form of $\mu r$ become negligible, whereas at large $r$, $Q_5^2/r^2$ rapidly approaches zero, and we are left with the Schwarzschild solution.

In the next section, we will establish a relationship between $Q_5$ and the source mass $M$. Using the values thus obtained, we find that the Reissner-Nordstr\"om solution is, in fact, a degenerate solution in which $B(r)$ never vanishes, and which would normally include a naked singularity. However, at extremely low values of $r$, our numerical solution deviates from the Reissner-Nordstr\"om solution (see Figure~\ref{fig:AB}.)
\end{itemize}

The numerical solutions discussed here require that initial values be established for the functions involved. Specifically, we require, at some $r=\tilde{r}$, the values of $G(\tilde{r})$, $G'(\tilde{r})$, $\mu(\tilde{r})$, $\mu'(\tilde{r})$, $\omega(\tilde{r})$, $\omega'(\tilde{r})$, $A(\tilde{r})$, $B(\tilde{r})$, $B'(\tilde{r})$, $\phi_t(\tilde{r})$, and $\phi_t'(\tilde{r})$. Note that $A'(\tilde{r})$ is absent from this list; the second derivative of $A(r)$ does not appear in our set of equations.

The values of $A$, $B$, and $B'$ are fixed by the requirement that at large distance from a source, we must be able to mimic the Schwarzschild solution (albeit with a modified gravitational constant), and that at spatial infinity, the metric must be asymptotically Minkowskian. The vector field $\phi$ must also vanish at infinity, which provides another boundary condition. Next, we assume that the values of $G$, $\mu$, and $\omega$ are dependent on the source mass only, i.e., $G'=\mu'=\omega'=0$. In total, we have established seven initial values or boundary conditions through these assumptions.

We seek the remaining initial conditions in the form of the fifth force charge $Q_5$, and initial values of $G=G_0$, $\mu=\mu_0$, and $\omega=\omega_0$. We note that the basic properties of the numerical solution and the solution's stability are not affected by the values chosen for these parameters. However, their values must be chosen such that they correctly reflect specific physical situations. To determine these values, we now turn to the case of the point test particle.

\section{Test particle equation of motion}
\label{sec:testpart}

We begin by defining a test particle via its Lagrangian:
\begin{equation}
{\cal L}_\mathrm{TP}=-m+\alpha\omega q_5\phi_\mu u^\mu,
\label{eq:TP}
\end{equation}
where $m$ is the test particle mass, $\alpha$ is a factor representing the nonlinearity of the theory (to be determined later), $\omega$ is present as it determines the interaction strength, $q_5$ is the test particle's fifth-force charge, and $u^\mu=dx^\mu/ds$ is its four-velocity.

We assume that the test particle charge is proportional to its mass:
\begin{equation}
q_5=\kappa m,
\label{eq:kappa}
\end{equation}
with $\kappa$ constant and independent of $m$. This assumption implies that the fifth force charge $q_5$ is not conserved, as mass is not conserved. This is the case in Maxwell-Proca theory, as $\nabla^\mu J_\mu\ne 0$.

From (\ref{eq:TP}), the equation of motion is obtained \cite{Moffat2006a,Moffat2008a}:
\begin{equation}
m\left(\frac{du^\mu}{ds}+\Gamma^\mu_{\alpha\beta}u^\alpha u^\beta\right)=-\alpha\kappa\omega mB^\mu{}_\nu u^\nu.
\end{equation}
That $m$ cancels out of this equation is nothing less than a manifestation of the equivalence principle.

Our numerical analysis of the spherically symmetric, static solution justifies the use of the Schwarzschild metric, provided that $r\gg 4\pi\omega G_0Q_5^2$. Further, for nonrelativistic motion, we can make the slow motion approximation $ds\simeq dt$, yielding the equation
\begin{equation}
\ddot{r}-\frac{J_N^2}{r^3}+\frac{GM}{r^2}=\alpha\kappa\omega\phi_t',
\end{equation}
where $\dot{r}=dr/dt$ and $J_N$ is the Newtonian angular momentum per unit mass. The second term on the left-hand side can be dropped at large $r$. Replacing $\phi_t'$ with the Yukawa solution (\ref{eq:phit}), and using (\ref{eq:kappa}) such that $Q_5=\kappa M$, we get
\begin{equation}
\ddot{r}=-\frac{GM}{r^2}\left[1-\frac{\alpha\kappa^2\omega}{G}(1+\mu r)e^{-\mu r}\right].
\label{eq:MOGaccel2}
\end{equation}
Compatibility with Newton's equation of motion requires that when $r\ll\mu^{-1}$, $\ddot{r}\simeq-G_NM/r^2$. Therefore,
\begin{equation}
G\left(1-\frac{\alpha\kappa^2\omega}{G}\right)=G_N,
\end{equation}
from which
\begin{equation}
\kappa^2=\frac{G-G_N}{\alpha\omega}.
\end{equation}
In particular, if we choose $\alpha$ such that
\begin{equation}
G=(1+\alpha)G_N,
\label{eq:Galpha}
\end{equation}
we get
\begin{equation}
\kappa=\sqrt{\frac{G_N}{\omega}}.
\end{equation}
Given $Q_5=\kappa M$, this allows us to write (\ref{eq:B}) in the form
\begin{equation}
B(r)=1-\frac{2G_NM}{r}+\frac{G_0G_NM^2}{r^2}.
\label{eq:BB}
\end{equation}
Further, we also get
\begin{equation}
Q_5\kappa\omega=G_NM.
\label{eq:bko}
\end{equation}
Having obtained these results, we can now use the test particle to probe the spherically symmetric field of a point-like source. First, we note that for large $r$, $A\simeq B\simeq 1$, and the equations (\ref{eq:mu},\ref{eq:omega},\ref{eq:G}) for the scalar fields $\mu$, $\omega$, and $G$ can be written as
\begin{equation}
\mu''+\frac{2}{r}\mu'-\frac{\mu'^2}{\mu}-\frac{G'}{G}\mu'-\frac{1}{4\pi}\omega G\phi_t^2\mu^3=0,
\label{eq:mu0}
\end{equation}
\begin{equation}
\omega''+\frac{2}{r}\omega'-\frac{G'}{G}\omega'+\frac{G}{8\pi}\phi_t'^2+\frac{G\mu^2\phi_t^2}{8\pi}=0,
\label{eq:omega0}
\end{equation}
\begin{equation}
G''+\frac{2}{r}G'-\frac{3}{2}\frac{G'^2}{G}+\frac{1}{2}\left(\frac{\mu'^2}{\mu^2}-\omega'^2\right)G=0.
\label{eq:G0}
\end{equation}

Given $\mu'=\omega'=0$, (\ref{eq:G0}) admits a constant solution. Conversely, holding $G'=\phi_t'=0$, the equations for $\mu$ and $\omega$ read

\begin{equation}
\mu''+\frac{2}{r}\mu'-\frac{\mu'^2}{\mu}-\frac{1}{4\pi}\omega G\phi_t^2\mu^3=0,
\label{eq:mu1}
\end{equation}
\begin{equation}
\omega''+\frac{2}{r}\omega'+\frac{G\mu^2\phi_t^2}{8\pi}=0.
\label{eq:omega1}
\end{equation}

If the dimensionless quantity $G\phi_t^2$ is small (which is the case if $\phi_t$ is calculated using (\ref{eq:phit}) for sufficiently large $r$), these equations uncouple from one another and admit approximately constant solutions, consistent with our numerical findings.

We now choose to perturb these equations by introducing a constant radial component to the $\phi$ field, such that $\phi_t^2\rightarrow\phi_t^2-\phi_r^2<0$. For shorthand, we write
\begin{equation}
\phi^2=\phi_r^2-\phi_t^2.
\end{equation}
Still keeping the equations uncoupled, they are now solved by the following expressions:
\begin{eqnarray}
\mu&=&\sqrt{\frac{4\pi}{\omega G\phi^2r^2}},\label{eq:mu0sol}\\
\omega&=&\frac{1}{48\pi}G\phi^2\mu^2r^2,\label{eq:omega0sol}\\
G&=&G_\infty\frac{r^2}{(r+C_1)^2}.\label{eq:G0sol}
\end{eqnarray}
How can these solutions be consistent with one another, and with our numerical results? First, we note that (\ref{eq:mu0sol}) and (\ref{eq:omega0sol}) define a constant value for $\omega$:
\begin{equation}
\omega_0=\frac{1}{\sqrt{12}}.
\end{equation}

Fixing $\omega$ in (\ref{eq:omega0sol}) also fixes $r$. We denote this value by $\tilde{r}$ and note that it can be expressed as a function of $M$:
\begin{equation}
\tilde{r}=K\sqrt{M},
\label{eq:r}
\end{equation}
with $K$ being a function of our chosen value of $\phi$:
\begin{equation}
K=\sqrt{\frac{48\pi G_N}{\kappa\mu^2Q_5G\phi^2}}.
\label{eq:K}
\end{equation}
{\em We choose $\phi_r$ to ensure that $K$ remains constant, independent of $M$.}
Putting this into (\ref{eq:mu0sol}) we obtain
\begin{equation}
\mu_0=\frac{D}{\sqrt{M}},
\label{eq:mu2}
\end{equation}
with $D$ given by
\begin{equation}
D=\sqrt{\frac{4\pi}{K^2\phi^2\omega G}}.
\end{equation}
Given $\mu\propto M^{-1/2}$, the value of $\mu^2Q_5$ is constant, since $Q_5\propto M$. Given constant $K$ in (\ref{eq:K}), this means that the dimensionless quantity $G\phi^2$ must also remain constant; this, in turn, means a constant $D$.

Conversely, from (\ref{eq:G0sol}) and (\ref{eq:r}), we get
\begin{equation}
G\equiv G_0=G_\infty\frac{M}{(\sqrt{M}+E)^2},
\label{eq:G2}
\end{equation}
with $E$ given by
\begin{equation}
E=\frac{C_1}{K}.
\end{equation}
These two constants $D$ and $E$ are universal; their values are independent of $M$. Having established their constancy, we can choose to determine $D$ and $E$ directly from observation (This, , in turn, would fix the values of $\tilde{r}$ and $\phi_r$, although it is not, in fact, necessary to calculate explicitly the values of these quantities.)

This result was obtained using a spherically symmetric, flat metric. Near a source, the flat metric is no longer a valid approximation; redoing the calculation using the Schwarzschild metric, we obtain
\begin{equation}
G\equiv G_0=G^*+(G_\infty-G^*)\frac{M}{(\sqrt{M}+E)^2},
\label{eq:G3}
\end{equation}
where agreement with Newtonian gravity demands $G^*=G_N$.

These equations represent our results for the initial values of $G$, $\omega$, and $\mu$ in our numerical solution: the numerical values of $D$ and $E$, in turn, along with $G_\infty$, can be determined from observation.

Our acceleration law (\ref{eq:MOGaccel2}), therefore, can be written as
\begin{equation}
\ddot{r}=-\frac{G_NM}{r^2}\left[1+\alpha-\alpha(1+\mu r)e^{-\mu r}\right],
\label{eq:Yukawa}
\end{equation}
with $\alpha$, using (\ref{eq:Galpha}) and (\ref{eq:G3}), given by
\begin{equation}
\alpha=\frac{M}{(\sqrt{M}+E)^2}\left(\frac{G_\infty}{G_N}-1\right),
\label{eq:alpha2}
\end{equation}
and $\mu=\mu_0$ given by (\ref{eq:mu2}).

The acceleration law (\ref{eq:Yukawa}) can also be recast in the commonly used Yukawa form:
\begin{equation}
\ddot{r}=-\frac{G_YM}{r^2}\left[1+\alpha_Y\left(1+\frac{r}{\lambda}\right)e^{-r/\lambda}\right],
\end{equation}
with the Yukawa parameters $\alpha_Y$ and $\lambda$ given by
\begin{eqnarray}
G_Y&=&\frac{G_N}{1+\alpha_Y},\\
\alpha_Y&=&-\frac{(G_\infty-G_N)M}{(G_\infty-G_N)M+G_N(\sqrt{M}+E)^2},\\
\lambda&=&1/\mu=\frac{\sqrt{M}}{D}.
\end{eqnarray}

We can also express the acceleration law (\ref{eq:Yukawa}) as
\begin{equation}
\ddot{r}=-\frac{G_\mathrm{eff}M}{r^2},
\end{equation}
where the effective gravitational constant $G_\mathrm{eff}$ is defined as
\begin{equation}
G_\mathrm{eff}=G_N\left[1+\alpha-\alpha(1+\mu r)e^{-\mu r}\right].
\label{eq:Geff}
\end{equation}

The metric parameter $B(r)$ can finally be written as
\begin{equation}
B(r)=1-\frac{2G_NM}{r}+\frac{(1+\alpha)G_N^2M^2}{r^2}.
\end{equation}
This is the solution that was shown in Figure~\ref{fig:AB}.

\section{Cosmology}
\label{sec:cosmo}

In the case of a homogeneous, isotropic cosmology, using the Friedmann-Lema\^itre-Robertson-Walker (FLRW) line element,
\begin{equation}
ds^2=dt^2-a^2(t)[(1-kr^2)^{-1}dr^2+r^2d\Omega^2],
\end{equation}
the field equations assume the following form:
\begin{equation}
\ddot{\mu}+3H\dot{\mu}-\frac{\dot{\mu}^2}{\mu}-\frac{\dot{G}}{G}\dot{\mu}+\frac{1}{4\pi}G\omega\mu^3\phi_0^2+\frac{2}{\mu}V_\mu-V'_\mu=0,
\end{equation}
\begin{equation}
\ddot{\omega}+3H\dot{\omega}-\frac{\dot{G}}{G}\dot{\omega}-\frac{1}{8\pi}G\mu^2\phi_0^2+\frac{1}{4\pi}GV_\phi+V'_\omega=0,
\end{equation}
\begin{eqnarray}
\ddot{G}+3H\dot{G}-\frac{3}{2}\frac{\dot{G}^2}{G}+\frac{G}{2}\left(\frac{\dot{\mu}^2}{\mu^2}-\dot{\omega}^2\right)+\frac{3}{G}V_G-V_G'+G\left[\frac{V_\mu}{\mu^2}+V_\omega\right]
\nonumber\\
+\frac{G}{8\pi}\Lambda-\frac{3G}{8\pi}\left(\frac{\ddot{a}}{a}+H^2\right)=0,
\end{eqnarray}
\begin{eqnarray}
H^2+\frac{k}{a^2}=\frac{8\pi G\rho}{3}
-\frac{4\pi}{3}\left(\frac{\dot{G}^2}{G^2}+\frac{\dot{\mu}^2}{\mu^2}-\dot{\omega}^2-\frac{1}{4\pi}G\omega\mu^2\phi_0^2\right)
\nonumber\\
+\frac{2}{3}
\omega GV_\phi+\frac{8\pi}{3}\left(\frac{V_G}{G^2}+\frac{V_\mu}{\mu^2}+V_\omega
\right)
+\frac{\Lambda}{3}+H\frac{\dot{G}}{G},
\label{eq:FR1}
\end{eqnarray}
\begin{eqnarray}
\frac{\ddot{a}}{a}=-\frac{4\pi G}{3}(\rho+3p)
+\frac{8\pi}{3}\left(\frac{\dot{G}^2}{G^2}+\frac{\dot{\mu}^2}{\mu^2}-\dot{\omega}^2-\frac{1}{4\pi}G\omega\mu^2\phi_0^2\right)
\nonumber\\
+\frac{2}{3}
\omega GV_\phi+\frac{8\pi}{3}\left(\frac{V_G}{G^2}+\frac{V_\mu}{\mu^2}+V_\omega
\right)
+\frac{\Lambda}{3}+H\frac{\dot{G}}{2G}+\frac{\ddot{G}}{2G}-\frac{\dot{G}^2}{G^2},
\label{eq:FR2}
\end{eqnarray}
\begin{equation}
\omega\mu^2\phi_0-\omega\frac{\partial V_\phi}{\partial\phi_0}=4\pi J_0,~~~~~~~~~~0=J_i,
\end{equation}
where $H=\dot{a}/a$ is the Hubble expansion rate.

\begin{figure}[t]
\hskip 0.2\linewidth\includegraphics[width=0.8\linewidth]{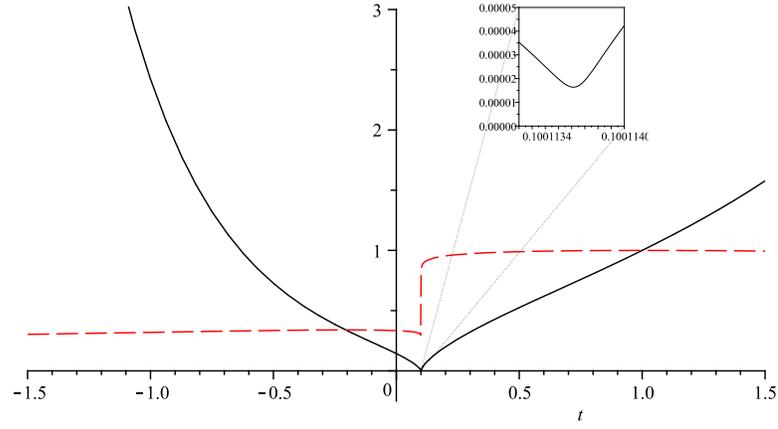}
\caption{The MOG ``bouncing'' cosmology. The horizontal axis represents time, measured in Hubble units of $H_0^{-1}$. The solid (black) line is $a/a_0$, the scale factor normalized to the present epoch. The dashed (red) line is $G/G_0$. The inset shows details of the bounce, demonstrating that a smooth bounce occurs even as the matter density of the universe is more than $10^{14}$ times its present value.}
\label{fig:aG}
\end{figure}

\begin{figure}[t]
\hskip 0.2\linewidth\includegraphics[width=0.8\linewidth]{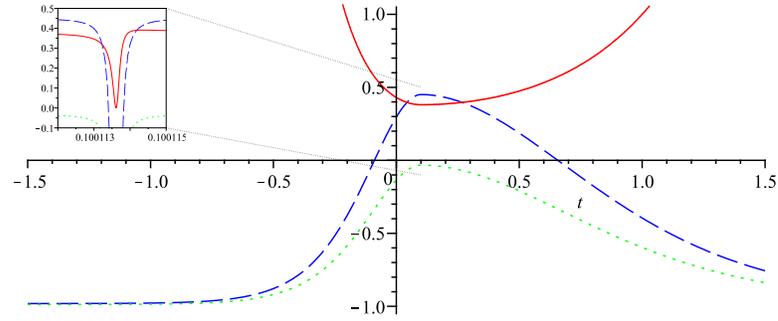}
\caption{The effective equation of state $w_\mathrm{eff}$ (dotted green line) and the deceleration parameter $q$ (dashed blue line) in MOG cosmology. Inset shows $q$ in the vicinity of the bounce. The behavior of $w$ is similar, in accordance with $q=(1+3w)/2$. Also shown with a solid red line is the effective density normalized to the cosmic scale parameter, computed as $(G_0 a(t)^3\rho_\mathrm{eff})/(G_N a_0^3\rho_0)$.}
\label{fig:qw}
\end{figure}

Just as we were able to do so in the spherically symmetric case, it is possible to obtain an exact numerical solution to this set of equations using numerical methods. While noting that the stability or qualitative properties of the solution are not sensitive to specific parameter values, for consistency with cosmological physics, we choose the following initial conditions:
\begin{eqnarray}
t_0&=&13.7\times 10^9~\mathrm{years},\\
a_0&=&ct_0,\\
G_0&=&6G_N,\\
\mu_0&=&H_0,\\
\omega_0&=&1/\sqrt{12},\\
\phi_0&=&0,\\
\dot{a}_0&=&H_0a_0=72a_0~\mathrm{km}/\mathrm{Mpc}~\mathrm{s},\\
\dot{G}_0=\dot{\mu}_0=\dot{\omega}_0&=&0,\\
V_G&=&0.07659537G_0^2/t_0^2,\\
V_\mu=V_\omega=V_\phi&=&0,\\
\Lambda&=&0,\\
k&=&0,
\end{eqnarray}

The values of $H_0$ and the Hubble time $t_0$, as well as that of $\dot{a}$, are the same as in standard cosmology. We used for $\omega_0$ the result from the previous section. For $\mu_0$, we simply assumed that it is the inverse of the Hubble scale $a_0$, which we believe to be a reasonable choice.

More notable is our choice of $G_0$. The value of $6G_N$ is motivated by two considerations. First, we note that given a baryonic matter content of $\Omega_b=\rho_b/\rho_\mathrm{crit}\simeq 0.05$ (where $\rho_b$ is the baryonic matter density, and $\rho_\mathrm{crit}=3H_0^2/8\pi G$ is the critical density required for a flat universe with $k=0$), a gravitational constant six times that of Newton decreases $\rho_\mathrm{crit}$ by the same factor without altering $\rho_b$ and, therefore, boosts $\Omega_b$ to the value of $\sim 0.3$, consistent with dark matter observations. Second, we note that an effective gravitational constant of $G_\mathrm{eff}\simeq 6G_N$ at the Yukawa distance $r=\mu^{-1}$, after solving for $\alpha$ in (\ref{eq:Geff}), yields $\alpha\simeq 19$ and an effective gravitational constant of $G_\mathrm{eff}\simeq 20G_N$ at infinity. On superhorizon scales our solution is consistent with an Einstein-de Sitter cosmology with no dark matter or dark energy. (This would imply a vanishing $V_G$ on superhorizon scales.)

To carry out the solution, we assume a pressureless matter equation of state $w=p/\rho=0$. We find that this results in a ``bouncing'' cosmology. The bounce can be fine-tuned by choosing an appropriate value for $V_G$, which we have done. This ensures that the universe reaches sufficient density in order to form a surface of last scattering. At this point, the equation of state $w=0$ is obviously no longer valid. However, introducing a mixed equation of state that also incorporates radiation does not alter substantially the qualitative features of our model.

We emphasize that in our model, only baryonic matter is present, with a matter density of $\sim$5\% of the critical density. Nevertheless, the cosmology is flat, due in part to the increased value of the gravitational constant, and in part to the presence of the nonzero energy density associated with $V_G$.

The values of $\mu$ and $\omega$ remain constant, while $G$ has a different value prior to the bounce, undergoing a rapid (but smooth) change at the time of the bounce.

We can define an effective equation of state by rewriting (\ref{eq:FR1}) and (\ref{eq:FR2}) as follows:
\begin{eqnarray}
H^2+\frac{k}{a^2}&=&\frac{8\pi G\rho_\mathrm{eff}}{3},\\
\frac{\ddot{a}}{a}&=&-\frac{4\pi}{3}(1+3w_\mathrm{eff})G\rho_\mathrm{eff},
\end{eqnarray}
and solving for $w_\mathrm{eff}$. We can also define the deceleration parameter as
\begin{equation}
q=-\frac{\ddot{a}a}{\dot{a}^2}.
\end{equation}

We find that $w_\mathrm{eff}$ is confined between $-1\lesssim w_\mathrm{eff}\lesssim 0$; it reaches its maximum at the time of the bounce, going asymptotically to its minimum value at times both after and prior to the bounce. The deceleration parameter $q$ is negative in the distant past (indicating slowing contraction). It becomes minus infinity at the time of the bounce (corresponding to $\dot{a}=0$), and becomes negative again at the present epoch, indicating accelerating expansion (see Figure~\ref{fig:qw}).

That MOG admits a bouncing cosmology without exotic matter or quantum effects may appear surprising at first, but a form of this bouncing solution is already known. As the $\mu$ and $\omega$ fields remain constant in this solution, they can be eliminated. The remaining field equations are those of Brans-Dicke theory \cite{BD1962} with $\omega_\mathrm{BD}=-8\pi$. Past investigations (see, e.g., \cite{GFR1973}) show that, indeed, a bouncing cosmology in Brans-Dicke theory can be obtained when $\omega_\mathrm{BD}<-6$. However, standard Brans-Dicke theory lacks a $V_G$ potential, which is what gave us the ability to fine-tune the solution and achieve a high density universe at the time the bounce occurs.

Brans-Dicke theory is known to be in conflict with solar system observations, except when $\omega_\mathrm{BD}$ is very large. However, we note that even though our cosmology is similar to Brans-Dicke cosmology, our spherically symmetric solution is not a Brans-Dicke solution. Here, the dominant contribution arises as a result of the $\phi_\mu$ field, which is absent from Brans-Dicke theory.

\section{Observations}
\label{sec:obs}

\begin{figure}[t]
${}$\hskip 0.2\linewidth\includegraphics[width=0.38\linewidth]{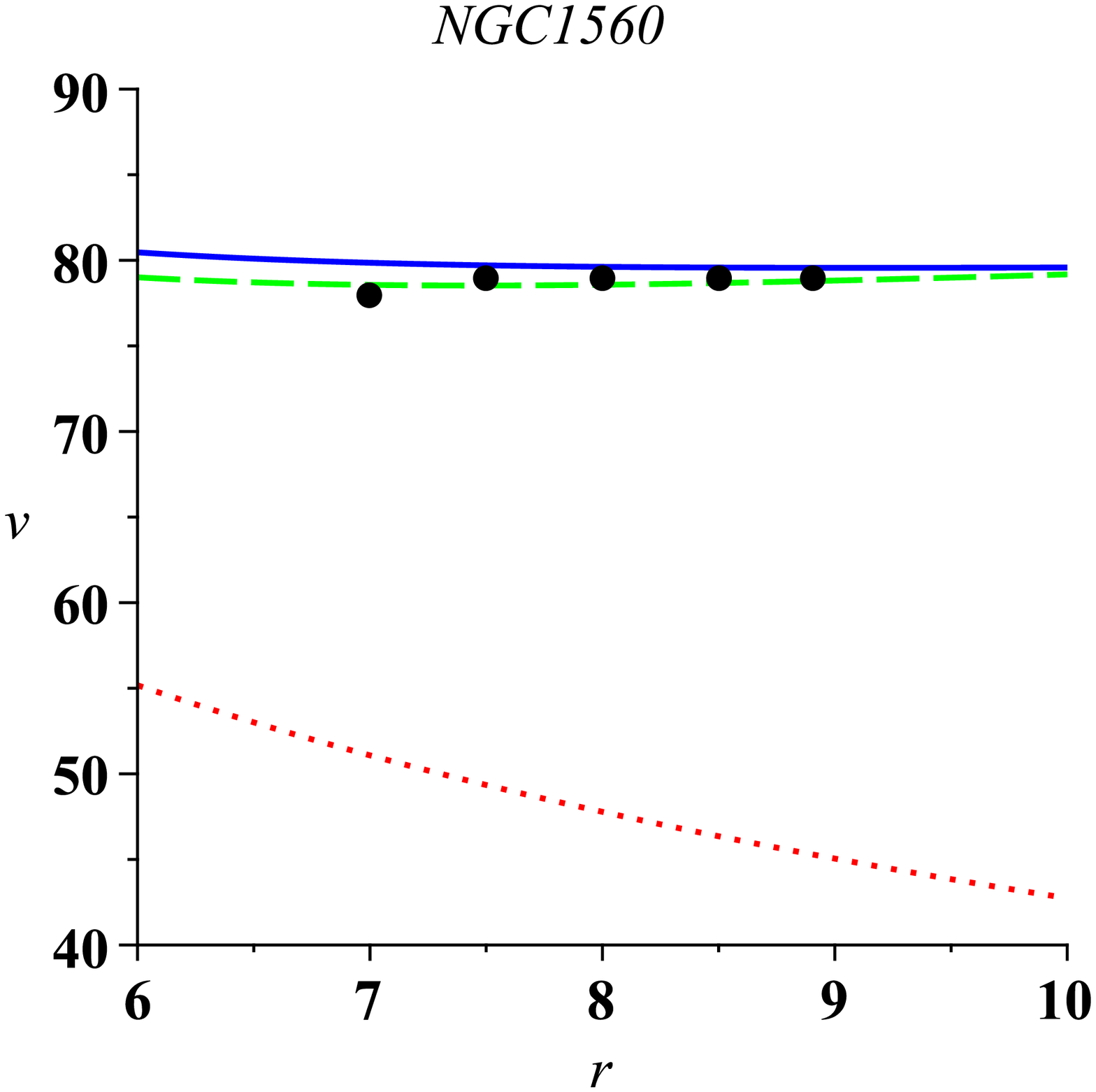}
\hskip 0.01\linewidth
\includegraphics[width=0.38\linewidth]{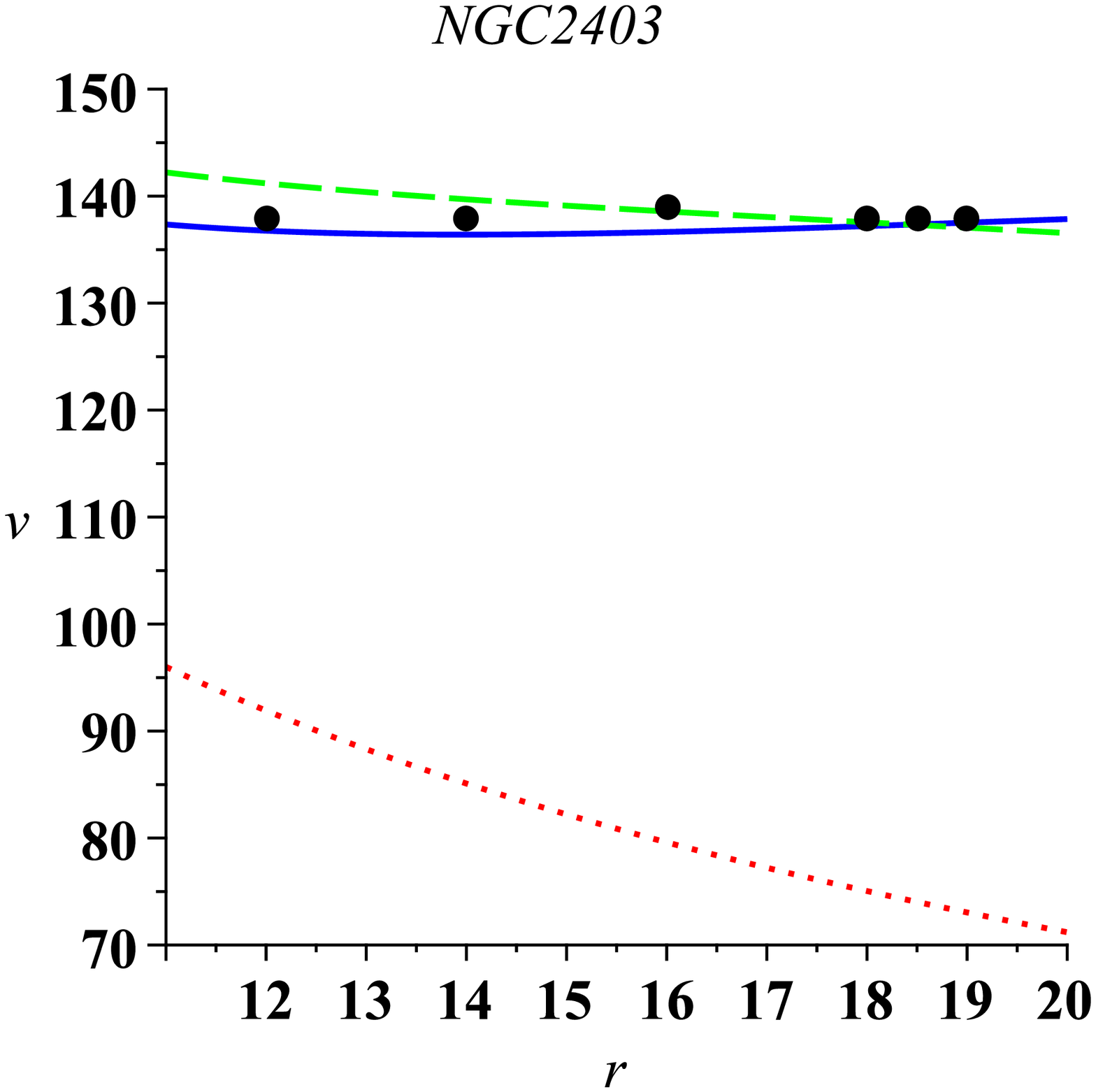}\\
${}$\hskip 0.2\linewidth\includegraphics[width=0.38\linewidth]{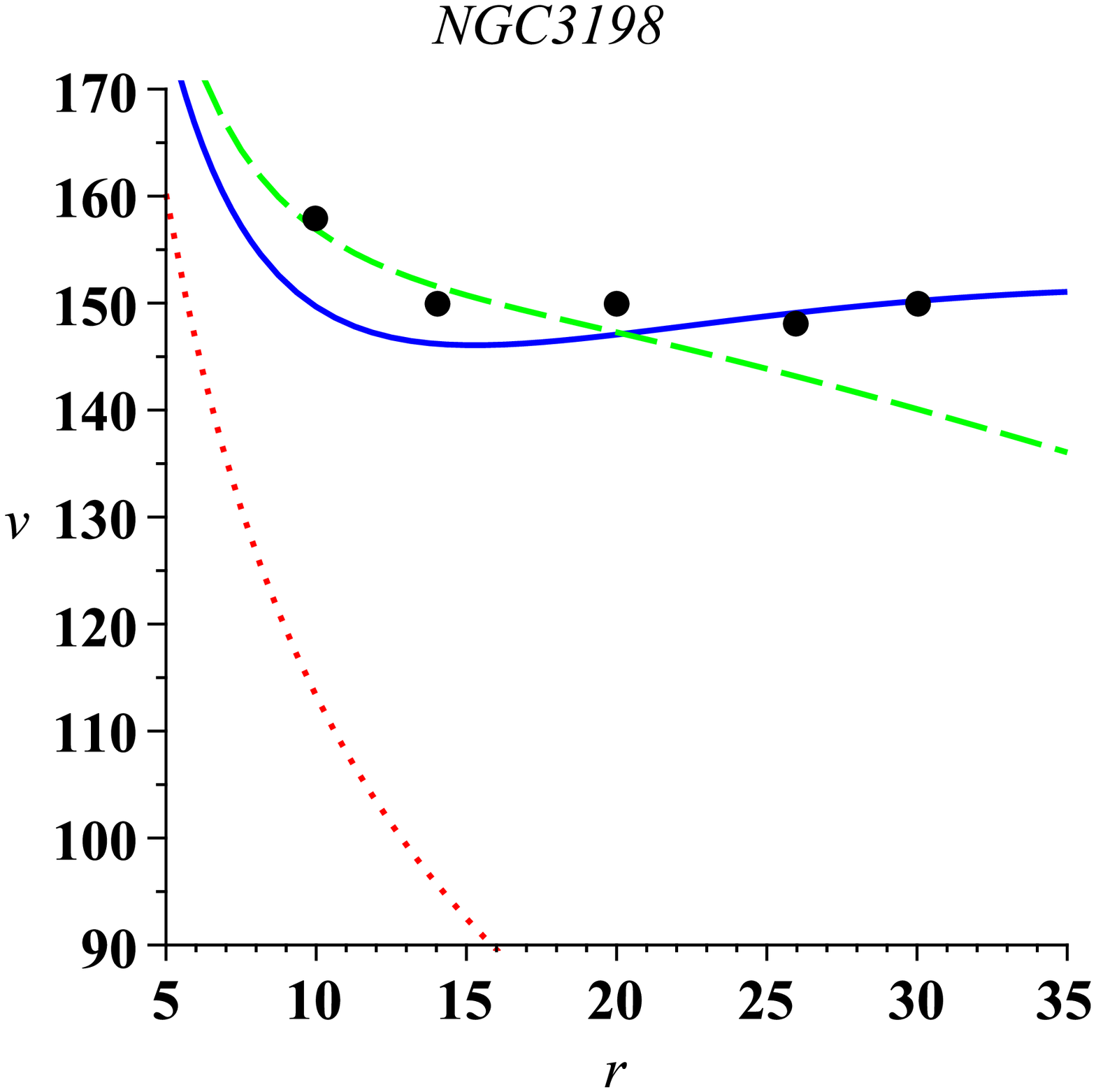}\hskip
0.01\linewidth
\includegraphics[width=0.38\linewidth]{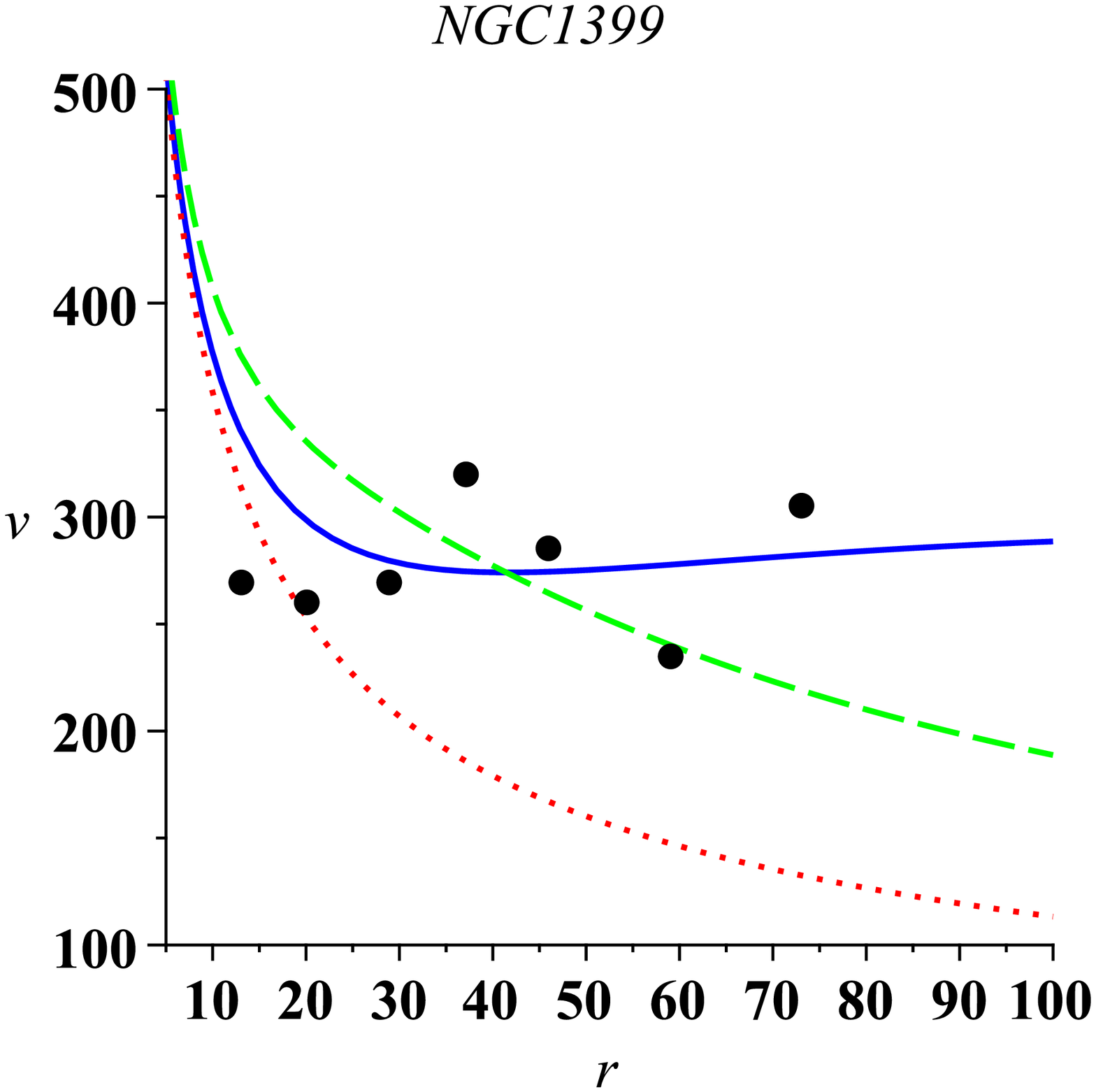}
\caption{Galaxy rotation curves for a small set of galaxies of varying size. Data points are marked as black dots, current rotational velocity estimates are represented by a solid (blue) curve, while the dashed (green) curve shows velocity estimates in accordance with our earlier work \cite{Moffat2004,Moffat2005}. Mass estimates are as in \cite{Moffat2004,Moffat2005}, except for NGC~1399, for which a mass estimate of $M=5\times 10^{11}~M_\odot$ was used. Dotted (red) curve is the Newtonian rotational velocity estimate for these galaxies using the same mass estimates. Radial distances are measured in kpc, masses in $M_\odot$.}
\label{fig:gals}
\end{figure}

Unless one assumes that a massive dark matter halo is present, a typical spiral galaxy is dominated in mass by the central bulge. The motion of stars in the outer reaches of a galaxy can, therefore, be well approximated by the equations of motion in a static, spherically symmetric vacuum field. Indeed, our experience shows that the flat rotation curves of galaxies provide a sensitive test to determine the values of the constants $D$ and $E$. In particular, it is easy to see that our results so far are compatible with the Tully-Fisher law \cite{TF1977}.

Kepler's laws of orbital motion yield a relationship between circular orbital velocity $v_c$ at radius $r$ from a mass $M$ in the form
\begin{equation}
\frac{v_c^2}{r}=\frac{GM}{r^2}.
\label{eq:vc}
\end{equation}

In contrast, Tully and Fisher determined that for most galaxies, assuming that the brightness of a galaxy and its mass are correlated, the flat part of the rotation curve obeys the empirical relationship
\begin{equation}
v_c^n\propto M,
\end{equation}
where $3\lesssim n\lesssim 4$.

In our case, from (\ref{eq:r}) and (\ref{eq:vc}) we obtain for $r\sim\tilde{r}$:
\begin{equation}
v_c^2\propto \frac{M}{\sqrt{M}}=\sqrt{M},
\end{equation}
corresponding to $n=4$ in the Tully-Fisher relationship.

Taking the next step, we have selected a small sample of galaxies that were studied earlier \cite{Moffat2004,Moffat2005}. An approximate fit to these galaxies yields the values
\begin{eqnarray}
D&\simeq&6250~M_\odot^{1/2}\mathrm{kpc}^{-1},\label{eq:C2}\\
E&\simeq&25000~M_\odot^{1/2}.\label{eq:C1}
\end{eqnarray}

The galaxy rotation curves we obtain for galaxies of varying mass (including the recently studied galaxy NGC~1399 \cite{Richtler2007}) are in good agreement with these values (Figure~\ref{fig:gals}) without dark matter.

The galaxy rotation curves in Figure~\ref{fig:gals} were obtained modeling the galaxies as point masses, without benefiting from a core model, or the use of photometric data, as in the more extensive fit to galaxy rotation velocities \cite{Brownstein2006a}. Nevertheless, this exercise demonstrates that our newly established relationships between $M$, $\alpha$, and $\mu$ not only satisfy the Tully-Fisher relationship in principle, but also offer good agreement with actual observations.

\begin{figure}[t]
${}$\hskip 0.2\linewidth\includegraphics[width=0.38\linewidth]{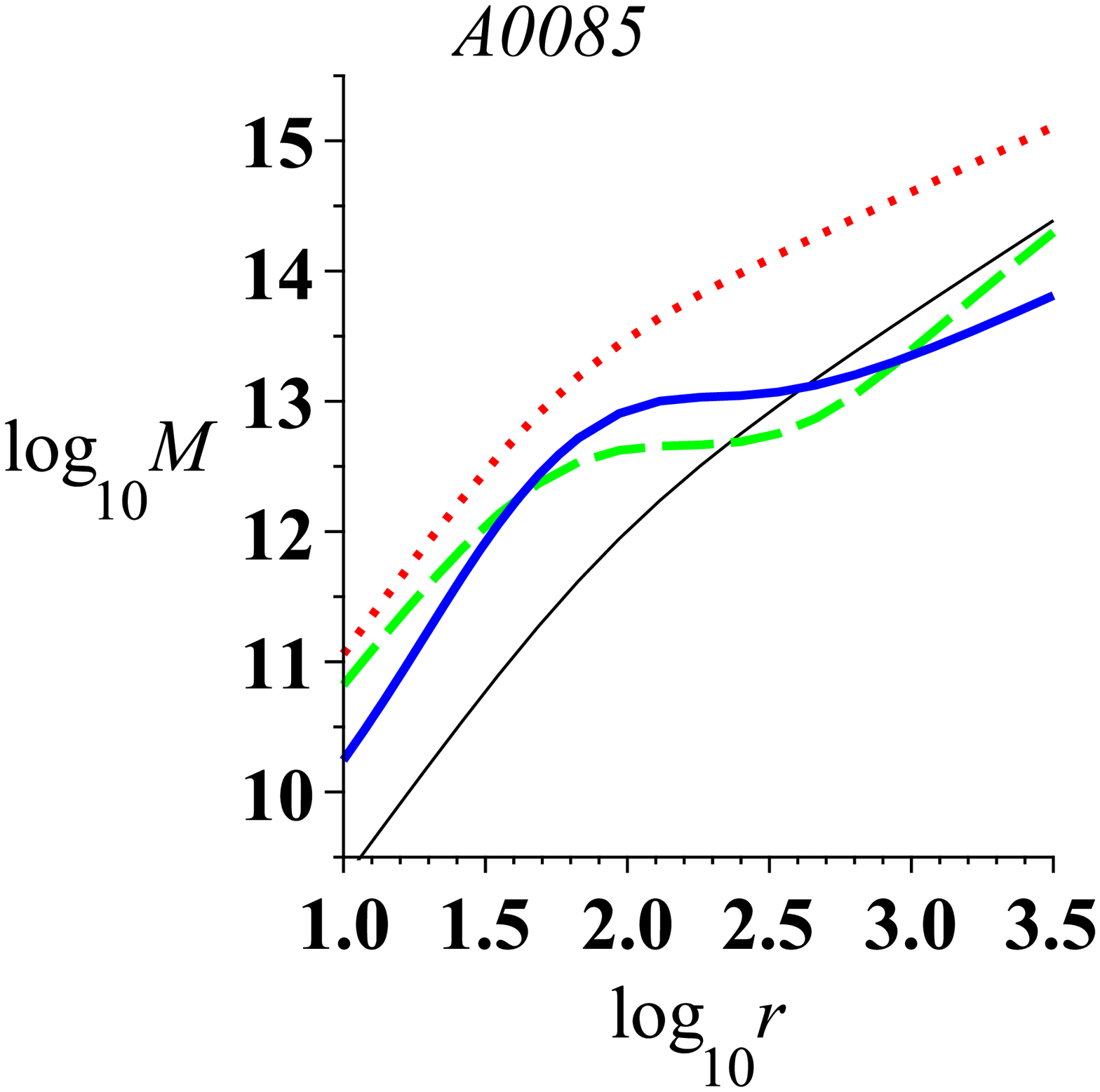}
\hskip 0.01\linewidth
\includegraphics[width=0.38\linewidth]{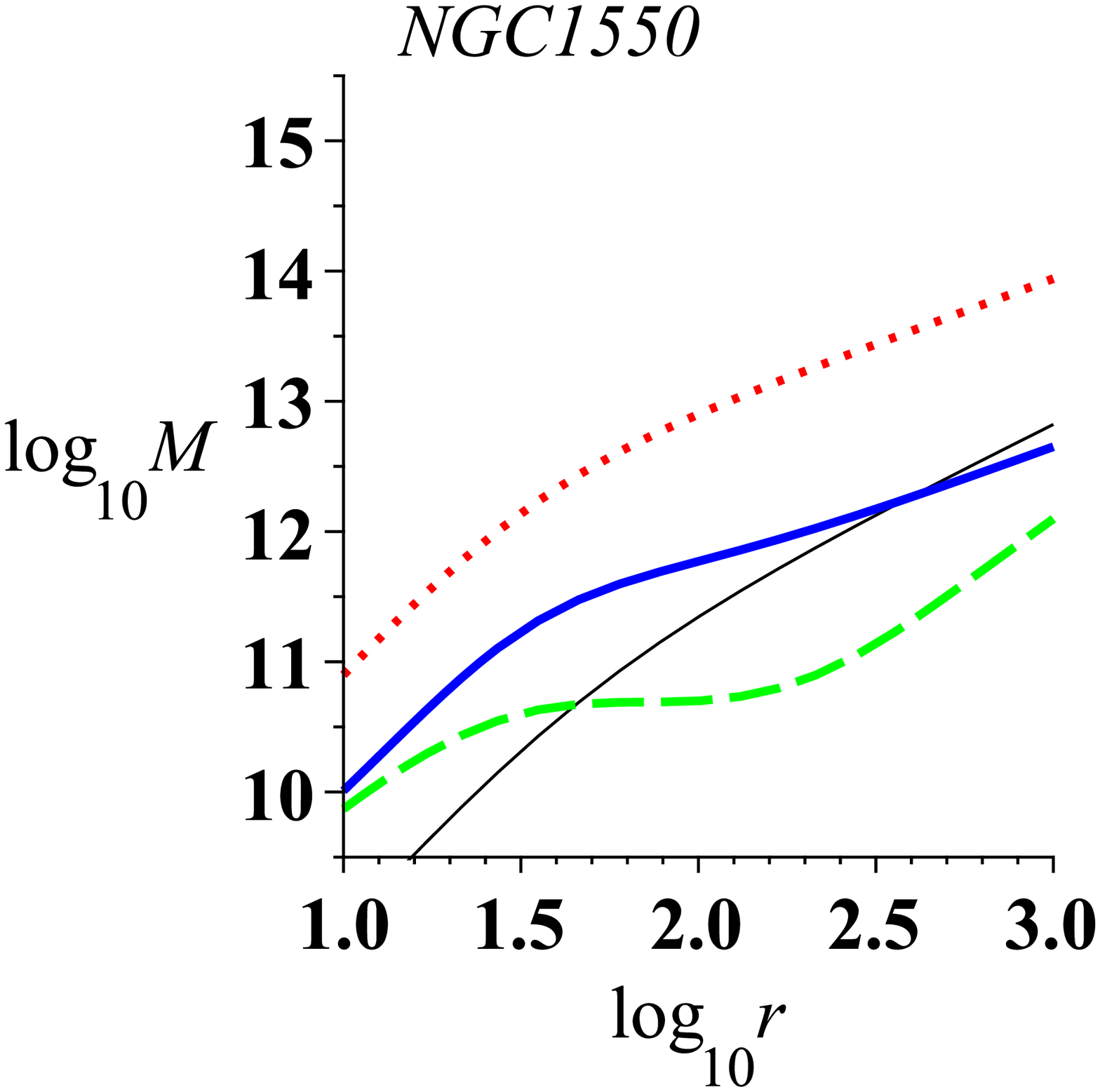}\\
${}$\hskip 0.2\linewidth\includegraphics[width=0.38\linewidth]{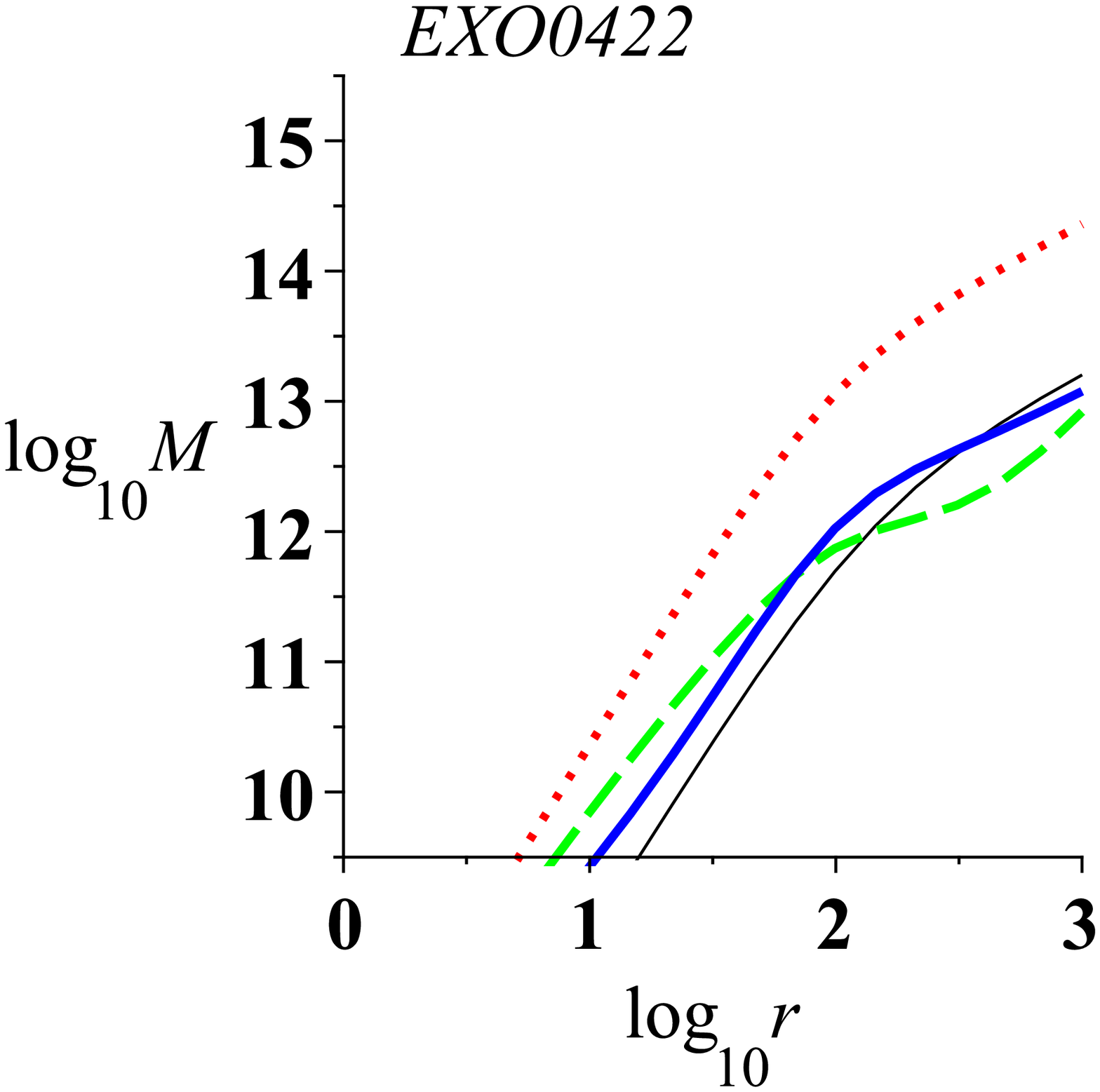}
\hskip 0.01\linewidth
\includegraphics[width=0.38\linewidth]{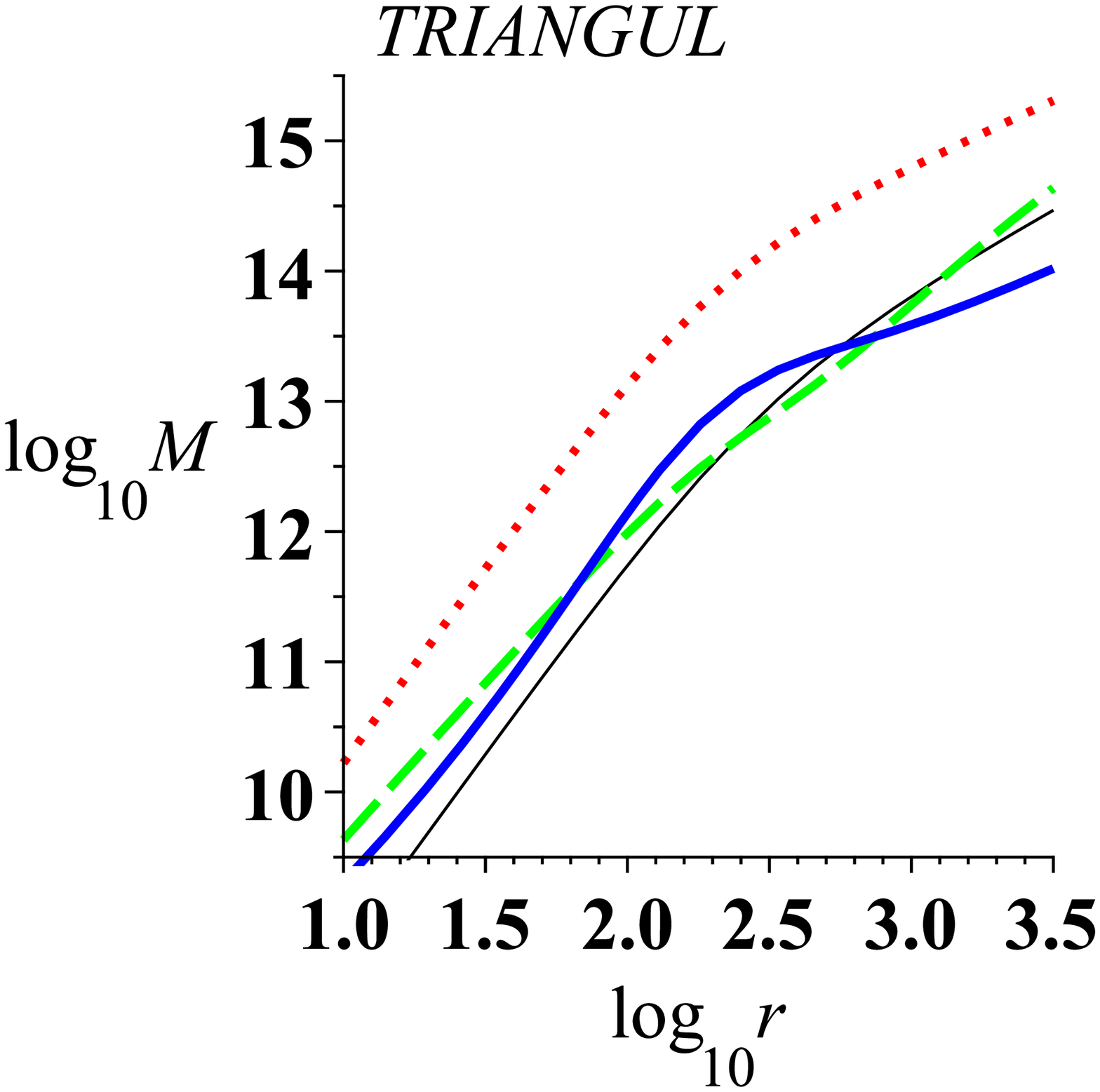}
\caption{A small sample of galaxy clusters studied in \cite{Brownstein2006b}. Thin (black) solid line is the mass profile estimate from \cite{RB2002}. Thick (blue) solid line is the mass profile estimated using our new results. Dashed (green) line is the result published in \cite{Brownstein2006b}, while the dotted (red) line is the Newtonian mass profile estimate. Radial distances are measured in kpc, masses in $M_\odot$.}
\label{fig:clusters}
\end{figure}

Is it a mere coincidence that we have obtained good agreement with the Tully-Fisher relationship, or do our solutions work on other scales? Using the already established values of $D$ and $E$, we attempted to reproduce the galaxy cluster mass profiles reported in \cite{Brownstein2006b}. Here, we run into a serious limitation: the solution we obtained is a spherically symmetric, static {\it vacuum} solution, and its applicability to an extended distribution of matter, such as that which one finds in a galaxy cluster, is limited. In a theory that offers linear behavior in the weak field limit, this is not necessarily a problem, as an extended distribution of matter can be well approximated by a large number of point masses. This is not so in our case: the gravitational field produced by two point sources is not simply the sum of their respective spherically symmetric static vacuum solutions.

This difficulty can only be addressed properly once appropriate interior solutions of our theory become known. Until then, our discussion is necessarily phenomenological: the best we can achieve is a parameterization of the difference between any result derived from the spherically symmetric, static vacuum solution and a physically relevant interior solution.

In \cite{Brownstein2006b}, the spherically symmetric, static vacuum solution was used successfully to model galaxy clusters. We are able to produce a comparable result while keeping the parameters $D$ and $E$ constant, by introducing an additional assumption: that the values of the MOG parameters $G_\infty$ and $\mu$ at some distance $r$ from the center of a spherically symmetric distribution of matter are determined not only by the amount of matter contained within radius $r$, but by the amount of matter within radius $r^*$. Figure~\ref{fig:clusters} shows the case of $r^*=3r$.

\begin{figure}[t]
${}$\hskip 0.2\linewidth\includegraphics[width=0.75\linewidth]{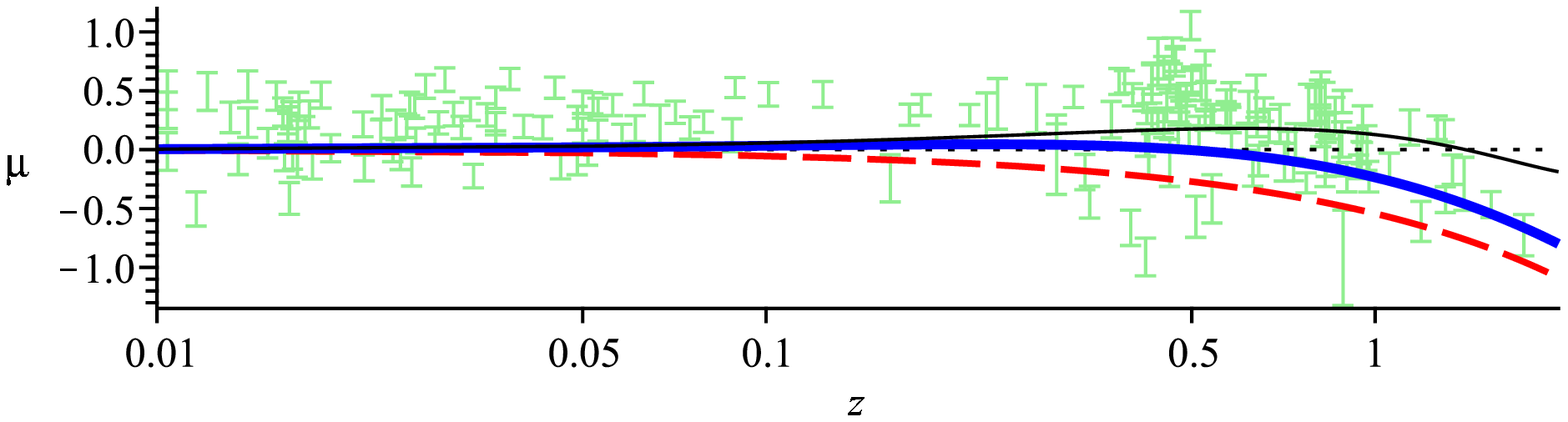}\\
${}$\hskip 0.2\linewidth\includegraphics[width=0.38\linewidth]{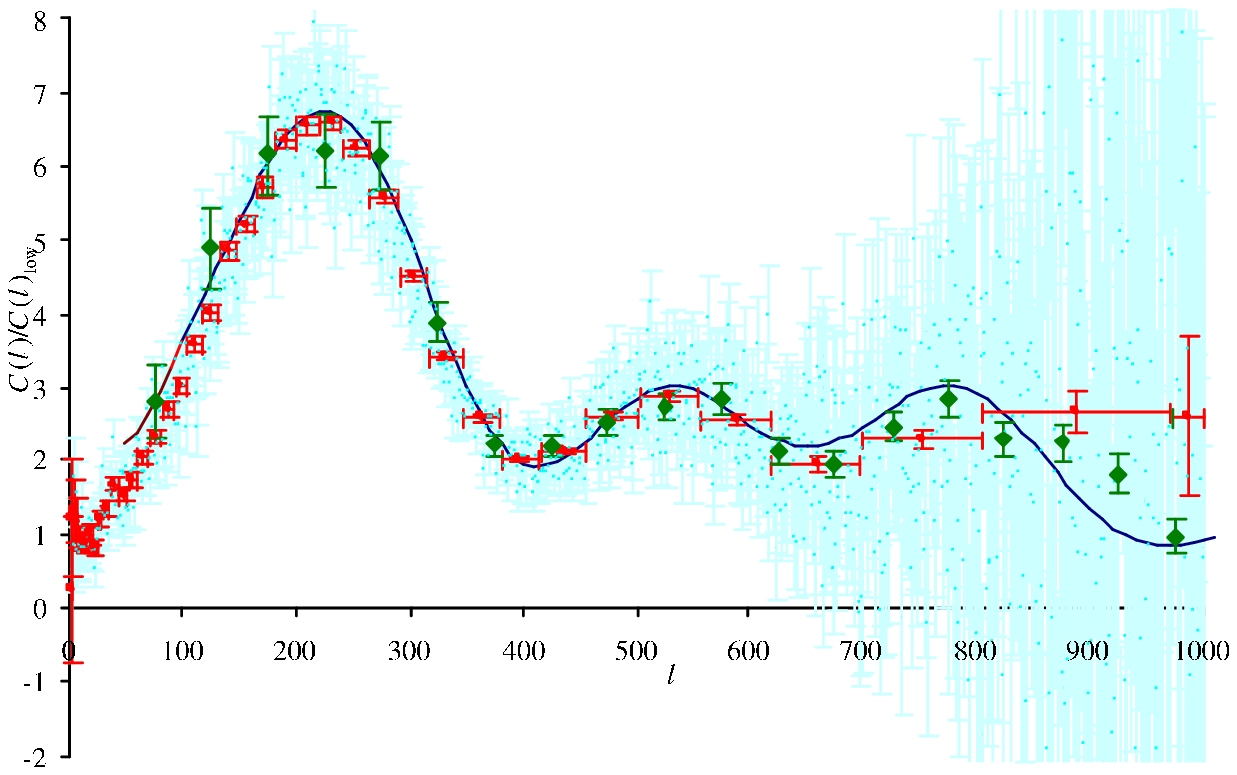}
\hskip 0.01\linewidth
\includegraphics[width=0.38\linewidth]{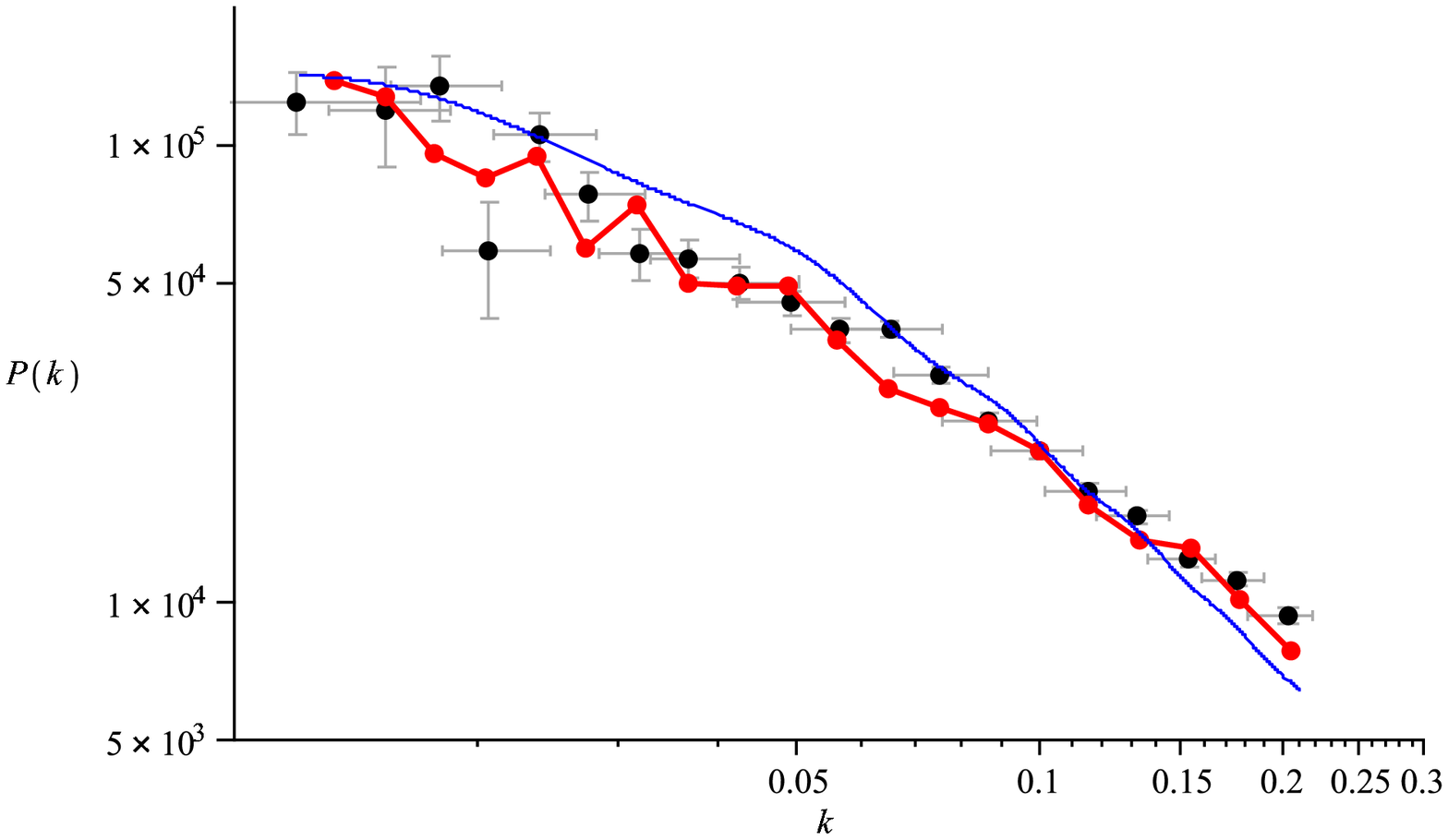}
\caption{Cosmological observations and Modified Gravity (from \cite{Moffat2007c}). Top panel: the luminosity-distance relationship of type Ia supernovae, with the MOG prediction shown with a thick (blue) line. (Thin (black) line is the $\Lambda$CDM prediction; dashed (red) line is a flat Einstein-de Sitter universe, while the horizontal access corresponds with an empty universe with no deceleration.) Bottom left: the angular CMB power spectrum showing good agreement between the MOG prediction and WMAP-3 and Boomerang data. Bottom right: After applying the appropriate window function, MOG (thick red line) shows agreement with the SDSS luminous red galaxy survey mass power spectrum, perhaps even superior to the $\Lambda$CDM prediction (thin blue line).}
\label{fig:cosmo}
\end{figure}

As we have shown \cite{Moffat2007c}, our theory also appears consistent with key cosmological observations, including the luminosity-distance relationship of type Ia supernovae, the angular CMB power spectrum, and the galaxy-galaxy mass power spectrum (Figure~\ref{fig:cosmo}).

If the acceleration law (\ref{eq:MOGaccel2}) along with (\ref{eq:mu2}) and (\ref{eq:alpha2}) is valid using the constants (\ref{eq:C2}) and (\ref{eq:C1}), it should offer agreement with other observations. In particular, the following expression must be constant:
\begin{equation}
\mu^2M=D^2.
\label{eq:Mr}
\end{equation}

\begin{figure}[t]
\hskip 0.2\linewidth\includegraphics[width=0.8\linewidth]{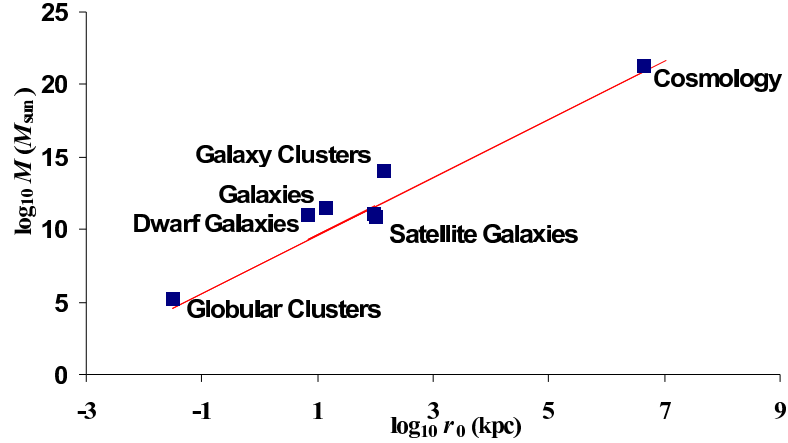}
\caption{The relationship $\mu^2M=$~const. between mass $M$ and the Yukawa-parameter $r_0=\mu^{-1}$ across many orders of magnitude remains valid. The solid red line represents our theoretical prediction in accordance with (\ref{eq:Mr}). We are using cosmological data from \cite{Moffat2007c}; galaxy cluster fits from \cite{Brownstein2006a}; galaxy and dwarf galaxy fits from \cite{Moffat2004,Moffat2005}; satellite galaxy fits from \cite{Moffat2007}; and globular cluster fits from \cite{Moffat2007a}. Note that the dwarf galaxy, galaxy, and galaxy cluster outliers are removed when these objects are recalculated using the results presented in this paper.}
\label{fig:Mr}
\end{figure}

To verify the validity of this relationship, we have plotted $M$ vs. $r_0=\mu^{-1}$ in Figure~\ref{fig:Mr}. For the purposes of this plot, we used previously published results, while noting that our new calculations place dwarf galaxies, galaxies, and galaxy clusters by definition exactly on the line representing our prediction. This plot demonstrates the validity of Eq.~(\ref{eq:Mr}) from the scales of star clusters to cosmological scales.

\begin{figure}[t]
\raggedleft
\includegraphics[width=0.48\linewidth,angle=270]{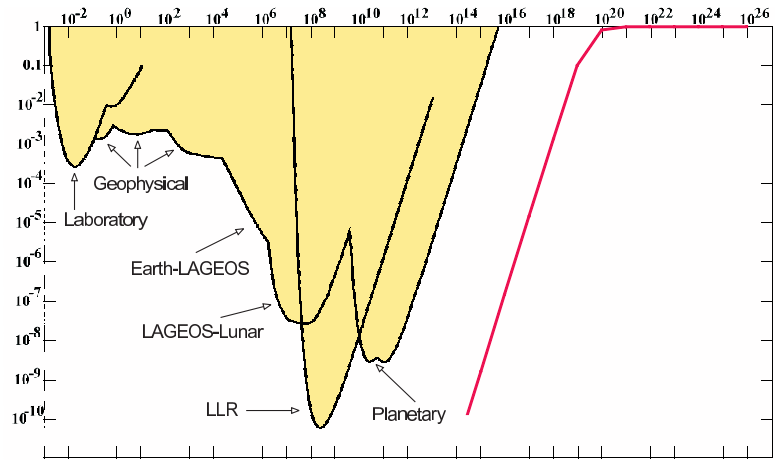}
\caption{Predictions of the Yukawa-parameters from the MOG field equations are not in violation of solar system and laboratory constraints. Predicted values of $\lambda$ (horizontal axis, in m) vs. $|\alpha_Y|$ are indicated by the solid red line. Plot adapted from \cite{Adelberger2003}.}
\label{fig:noses}
\end{figure}

The theory must also be consistent with observations or experiments performed within the solar system or in Earthbound laboratories. Several studies (see, e.g., \cite{Adelberger2003}) have placed stringent limits on Yukawa-like modifications of gravity based on planetary observations, radar and laser ranging, and other gravity experiments. However, our prediction of the absolute value of the $\alpha_Y$ parameter is very small when $\lambda_Y$ is small. The latter is estimated at $\lambda_Y\simeq 0.16$~pc ($\sim 5\times 10^{15}$~m, or about 33,000~AU) for the Sun, and $\lambda_Y\simeq 2.8\times 10^{-4}$~pc ($\sim 8.7\times 10^{12}$~m, or $\sim 58$~AU) for the Earth. The corresponding values of $|\alpha_Y|$ are $|\alpha_Y|\simeq 3\times 10^{-8}$ and $|\alpha_Y|\simeq 9\times 10^{-14}$, respectively, clearly not in contradiction with even the most accurate experiments to date (Figure~\ref{fig:noses}).

While these observations do not prove that MOG is correct, they do suggest that MOG might be falsifiable. On the scale of Earth-based laboratories and in the solar system, experiments with ever greater precision might eventually rule out MOG. Beyond the solar system, as larger galaxy samples become available, the presence or absence of baryonic oscillations in the matter power spectrum may unambiguously decide in favor of modified gravity theories or dark matter. Confirmed detection of dark matter particles in deep space or in the laboratory would also be a strong indication against modified gravity.

\section{Conclusions}
\label{sec:concl}

In this paper, we have demonstrated how results of our Modified Gravity theory can be derived directly from the action principle, without resorting to the use of fitted parameters. After we fix the values of some integration constants from observations, no free adjustable parameters remain, yet the theory remains consistent with observational data in the two cases that we examined: the vacuum solution of a static, spherically symmetric gravitational field, and a cosmological solution. These solutions were explored using numerical methods, avoiding the necessity to drop terms or make other simplifying assumptions in order to obtain an analytic solution. Further, the constraints used to compute the solutions are consistent to the extent that they overlap with one another, as discussed in Section~\ref{sec:cosmo}. We view these results as encouraging, albeit not conclusive. The fact that at the level of the calculations presented here, our theory is not obviously falsified is an indication that it may be a worthwhile effort to pursue MOG further, for instance by obtaining interior solutions to the MOG field equations, or by developing tools to perform $N$-body simulations to achieve more convincing results.

\section*{Acknowledgments}

We thank Joel Brownstein, Martin Green, and Pierre Savaria for helpful discussions.

The research was partially supported by National Research Council of Canada. Research at the Perimeter Institute for Theoretical Physics is supported by the Government of Canada through NSERC and by the Province of Ontario through the Ministry of Research and Innovation (MRI).

\section*{References}

\bibliography{refs}
\bibliographystyle{unsrt}

\end{document}